\documentclass[aps,prc,twocolumn,groupedaddress,showpacs]{revtex4}

\usepackage{graphicx}
\usepackage[rightcaption]{sidecap}
\usepackage{color}

\begin{document}
\title{Existence of inelastic supernumerary nuclear rainbow in   $^{16}$O+$^{12}$C  scattering
}

\author{S. Ohkubo$^{1}$,   Y. Hirabayashi$^2$ and A.  A. Ogloblin$^3$     
}
\affiliation{$^1$Research Center for Nuclear Physics, Osaka University, 
Ibaraki, Osaka 567-0047, Japan }
\affiliation{$^2$Information Initiative Center, Hokkaido University, Sapporo 060-0811, Japan}
 \affiliation{$^3$National Research Center  ``NRC Kurchatov Institute'',  RU-123182 Moscow, Russia}

\date{\today}

\begin{abstract}
\par
 The existence of a supernumerary nuclear rainbow in inelastic scattering is reported.
This is done by studying   inelastic $^{16}$O  scattering from $^{12}$C, exciting the  $2^+$ (4.44 MeV) state   
 of $^{12}$C and elastic scattering   at the  incident energies in the range  124 to  200 MeV, using 
 the coupled channels method.   An extended double folding  potential  is used. This is derived from  realistic  wave functions for 
  $^{12}$C  and $^{16}$O calculated with a microscopic  $\alpha$ cluster model and   a finite-range  density-dependent 
  nucleon-nucleon force.  Excitations to the  $2^+$ (4.44 MeV), 3$^-$ (9.64 MeV) and  $4^+$ (14.08 MeV) states of $^{12}$C, and  the $3^-$ (6.13 MeV) and $2^+$ (6.92 MeV) states of $^{16}$O are included in the coupled channels calculations. 
  The emergence of the supernumerary bow is understood by the properties of both the Luneburg-lens-like potential in
the internal region and diffuse attraction in the outer region. 
   The existence of a  supernumerary rainbow for inelastic scattering in addition to the existence of a dynamically created secondary rainbow and a  dynamically refracted primary rainbow  for elastic scattering, which are not observed in  meteorological rainbows,  further deepens  the understanding of   nuclear rainbows. 

 \end{abstract}

\maketitle

\par
\section{INTRODUCTION}
The existence of  supernumerary bows  in the inner bright side of the meteorological primary rainbow  was first explained by Airy in 1938    to be caused by the wave nature of light  \cite{Airy1938,Nussenzveig1977,Jackson1999,Adam2002}.
The supernumerary bows also appear in the outer bright side of its secondary rainbow.
 The  classical concept of the rainbow phenomenon was found to persist in quantum systems where the dual nature of particle and waves  dominates. Hundhausen and Pauly \cite{Hundhausen1965} made the first observation of an atomic rainbow associated  with the two supernumerary bows in the bright side of the primary rainbow for the  elastic scattering of  Na atoms from  Hg atoms.  The first observation of a nuclear rainbow was made by Goldberg {\it et al.}  for elastic $\alpha$ particle scattering from $^{58}$Ni \cite{Goldberg1974}. Nuclear rainbows have been widely observed in elastic $\alpha$ particle scattering and heavy-ion scattering under incomplete absorption, which  
 excludes a large class of potential ambiguities of   the nucleus-nucleus interaction potential  \cite{Brandan1997,Michel1998,Khoa2007}. The  potential that describes rainbow scattering has been powerful in the study of the  cluster structure of nuclei such as $\alpha$ cluster structure in $^{44}$Ti \cite{Michel1998} and the superdeformed $^{16}$O+$^{16}$O cluster structure in $^{32}$S \cite{Ohkubo2002,Ohkubo2003}.  The existence of supernumerary bows in nuclear  rainbow scattering, which is often called Airy structure after Airy,   have been observed  most clearly in heavy ion scattering  such as $^{16}$O+$^{16}$O,   $^{16}$O+$^{12}$C and $^{12}$C+$^{12}$C in the energy range between 5 and 10 MeV per nucleon  \cite{Khoa2007,Ogloblin1998,Nicoli1999,Nicoli2000,Ogloblin2000,Szilner2001,Ogloblin2003,Stokstad1979,Michel2004A}.

 \par
  Supernumerary bows in inelastic scattering,  which  are not expected in  meteorological rainbows but are possible in quantum systems,  have   been observed in molecular rotational rainbows such as   Na$_2$ scattering from Ne atoms  \cite{Hefter1981,Gottwald1987}. 
  The nuclear rainbow in inelastic scattering has  been  observed and studied extensively in Refs.%   
\cite{Bohlen1982,Bohlen1985,Brandan1986,Bohlen1993,Michel2004B,Michel2005,Khoa2005,Khoa2007,Ohkubo2014C}.
 The  existence of nuclear rainbows in inelastic   scattering makes it possible to  understand the interaction potential
 for the inelastic channels up to the internal region. In fact, the interaction potential determined in inelastic nuclear rainbow scattering   has been powerful in studying cluster structure. For example,  $\alpha$ particle condensation in the Hoyle state of $^{12}$C \cite{Ohkubo2004,Ohkubo2007,Belyaeva2010,Hamada2013}, four $\alpha$ cluster structure in $^{16}$O \cite{Ohkubo2010} and $\alpha$+$^{16}$O cluster structure with core excitation in $^{20}$Ne near the threshold energy region \cite{Hirabayashi2013}. 
 
 \par 
 However, the existence of a subtle supernumerary  bow in inelastic nuclear rainbow  scattering   has not been reported until now.  For the most typical  $^{16}$O+$^{16}$O system  the  difficulty of  resolution of the very close first excited  $0^+$ (6.05 MeV)  and the second excited $3^-$ (6.13 MeV) states of $^{16}$O    hampered the identification of the Airy minimum   for inelastic rainbow scattering \cite{Khoa2005}. On the other hand, for the $^{16}$O+$^{12}$C system there is no such  resolution problem for the first excited $2^+$ state of $^{12}$C at 4.44 MeV.  Also  the Airy minima in the angular distributions are not  obscured  by the  symmetrization, which occurs  for  systems  with two identical bosons  like $^{16}$O+$^{16}$O and   $^{12}$C+$^{12}$C. For   $^{16}$O+$^{12}$C there are  systematic experimental data of  elastic rainbow scattering  over a wide range of incident energies at $E_L=62-$1503 MeV
 \cite{Brandan1986,Ogloblin1998,Nicoli2000,Ogloblin2000,Szilner2001,Trzaska2002,Ogloblin2003} and we have studied the specific mechanism of nuclear rainbows for this system  such as  the existence of a dynamically generated secondary bow \cite{Ohkubo2014},  a ripple structure \cite{Ohkubo2014B} and a dynamically refracted primary rainbow \cite{Ohkubo2016}.  
  The present authors  were recently able to systematically  verify    the existence of the  Airy minimum, $A1$, of the primary rainbow  for  newly measured  $^{16}$O+$^{12}$C inelastic scattering in the energy range $E_L$=170-281 MeV \cite{Ohkubo2014C}.  A  systematic  evolution of the angular position of the  Airy minimum $A1$  with the  inverse of the center-of-mass  energy   was revealed.  We note that the inelastic $^{16}$O+$^{12}$C scattering measured in Ref.\cite{Szilner2006}  in the energy range   where the supernumerary bows appear in the elastic channel has not  been paid attention from the viewpoint of a supernumerary bow. It is intriguing and challenging to investigate the  existence of  supernumerary bows  in inelastic rainbow scattering.

\par
The purpose of this paper is to  report    the existence of  inelastic supernumerary nuclear  rainbows in    $^{16}$O+$^{12}$C scattering    in addtion to the $A1$ and  to study their  Airy structure through a coupled channels analysis of    the  inelastic and elastic angular distributions of differential cross sections.

\section{ EXTENDED FOLDING MODEL}
\par
We study    $^{16}$O+$^{12}$C  scattering  with the coupled channels   method  using 
an extended  double folding (EDF)  model that describes all the diagonal and off-diagonal
coupling potentials derived from  the microscopic   realistic wave functions for $^{12}$C  
and $^{16}$O  using  a density-dependent   nucleon-nucleon force.
  The diagonal and coupling potentials for the $^{16}$O+$^{12}$C system are calculated using
 the EDF  model.  We introduce the normalization factor $N_R$  \cite{Brandan1997} for the real double folding potential. 
\begin{eqnarray}
\lefteqn{V_{ij,kl}({\bf R}) = N_R
\int \rho_{ij}^{\rm (^{16}O)} ({\bf r}_{1})\;
     \rho_{kl}^{\rm (^{12}C)} ({\bf r}_{2})} \nonumber\\
&& \times v_{\it NN} (E,\rho,{\bf r}_{1} + {\bf R} - {\bf r}_{2})\;
{\it d}{\bf r}_{1} {\it d}{\bf r}_{2} ,
\end{eqnarray}
\noindent where $\rho_{ij}^{\rm (^{16}O)} ({\bf r})$ is the diagonal ($i=j$)  or transition ($i\neq j$)
 nucleon  density of  $^{16}$O    taken from  the microscopic $\alpha$+$^{12}$C  cluster model 
 wave functions calculated  with  the orthogonality condition model (OCM) from Ref.\cite{Okabe1995}. This model uses  a  realistic size parameter   for the $\alpha$ particle and for $^{12}$C  and is an extended version of  the  
OCM $\alpha$ cluster model  of Ref. \cite{Suzuki1976}, which reproduces  almost  all the energy levels  
well up  to $E_x$$\approx$13 MeV and the  electric transition probabilities   for $^{16}$O. 
We take into account  the important transition densities available in Ref.\cite{Okabe1995}, i.e.  g.s $\leftrightarrow$  $3^-$ (6.13 MeV) and $2^+$ (6.92 MeV)  in addition to all the  diagonal densities.  $\rho_{kl}^{\rm (^{12}C)} ({\bf r})$ represents the diagonal ($k=l$) or transition ($k\neq l$)  nucleon density of $^{12}$C  calculated using the microscopic three $\alpha$ cluster model of the resonating group method \cite{Kamimura1981}. This model reproduces the structure of  $^{12}$C  well  and the  wave functions     have  been checked against  many experimental  data,
 including charge form factors and electric transition probabilities \cite{Kamimura1981}. 
In the coupled channels calculations we  take into account  the   0$^+_1$ (0.0 MeV), $2^+$ (4.44 MeV),    3$^-$ (9.64 MeV), and  $4^+$ (14.08 MeV)  states of $^{12}$C.
 The mutual excitation channels in which both   $^{12}$C and $^{16}$O are excited simultaneously
 are not   included.  For the  effective interaction   $v_{\rm NN}$     we use  
 the DDM3Y-FR interaction \cite{Kobos1982}, which takes into account the
 finite-range    exchange effect \cite{Khoa1994}.
 An imaginary potential (non-deformed) is introduced   phenomenologically
 for all the diagonal potentials to take into  account the effect
of absorption due to other channels, which was successful in the recent coupled channels  studies of $^{16}$O+$^{12}$C 
 rainbow scattering   \cite{Ohkubo2014,Ohkubo2014B,Ohkubo2014C}.  Off-diagonals are
assumed to be  real.   Coulomb excitation is included.

%fig1
\begin{figure*}[t]
\includegraphics[keepaspectratio,angle=-90,width=17.4cm] {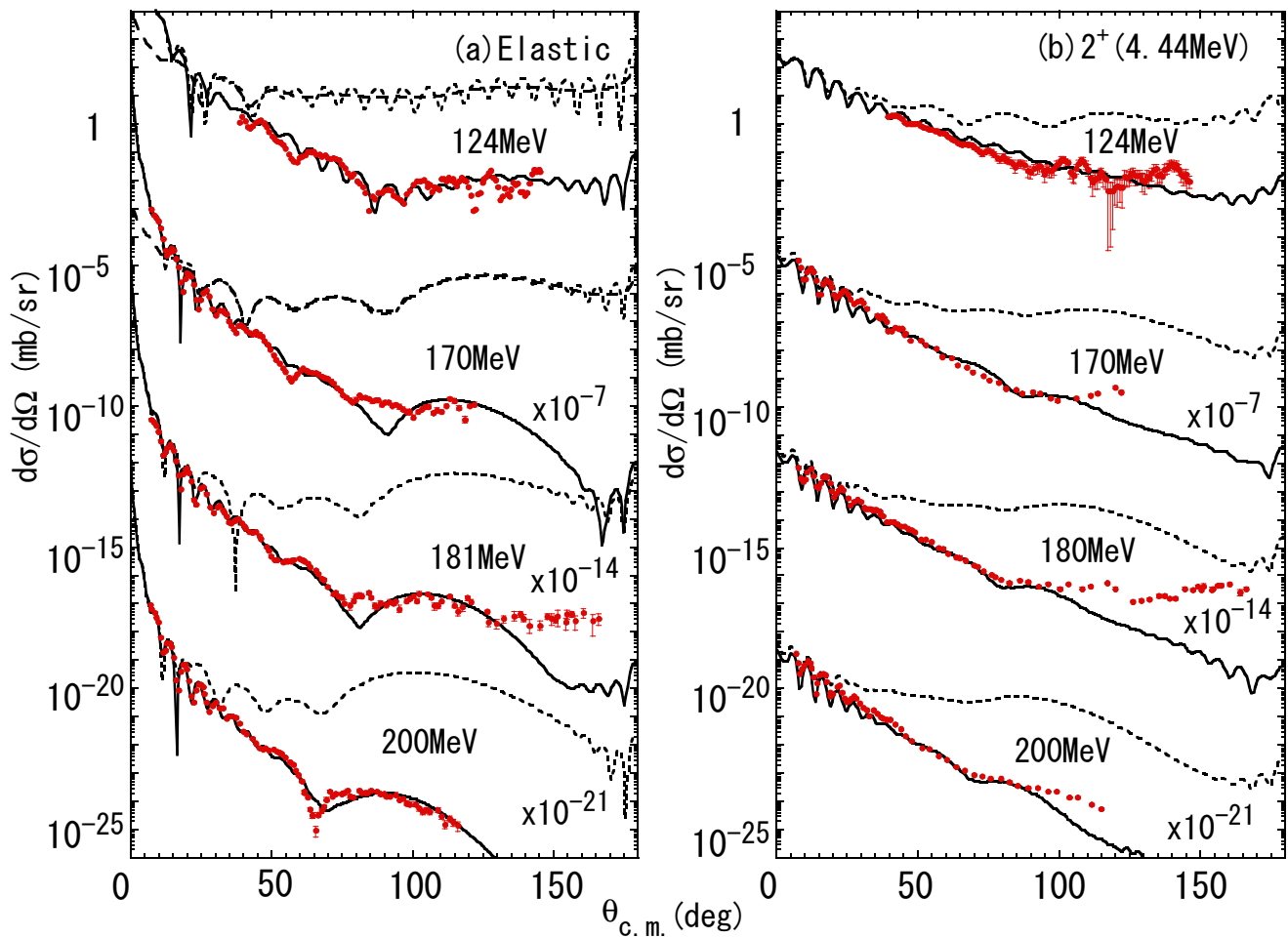}% angle=-90,,Here is how to import EPS art %%%
\protect\caption{\label{fig.1} {(Color online) 
Angular distributions in $^{16}$O+$^{12}$C  scattering at  $E_L$=124, 170, 181 (180)  and 200 MeV
calculated with the potentials in Table I  using the coupled channels method (solid lines) 
are compared with the experimental data (filled circles with vertical error bars) taken from \cite{Ogloblin2000,Szilner2006,Ohkubo2014C},
  (a)  elastic scattering   and  (b) inelastic scattering to the $2^+$ state  of  $^{12}$C.  
 The dotted lines  are calculated by switching off the imaginary  potentials in Table I in panel (a) and  only the imaginary potentials in the $2^+$ channel are switched off in panel (b).
 In panel (a) farside components calculated by switching off the imaginary potentials are displayed by with  dashed lines for $E_L=$ 124 and 170 MeV. 
 }
}
\end{figure*}

\section{COUPLED CHANNELS ANNALYSIS AND  SUPERNUMERARY BOW IN INELASTIC SCATTERING}
%fig2
\begin{figure*}[t]
 \includegraphics[keepaspectratio,angle=-90,width=14.0cm] {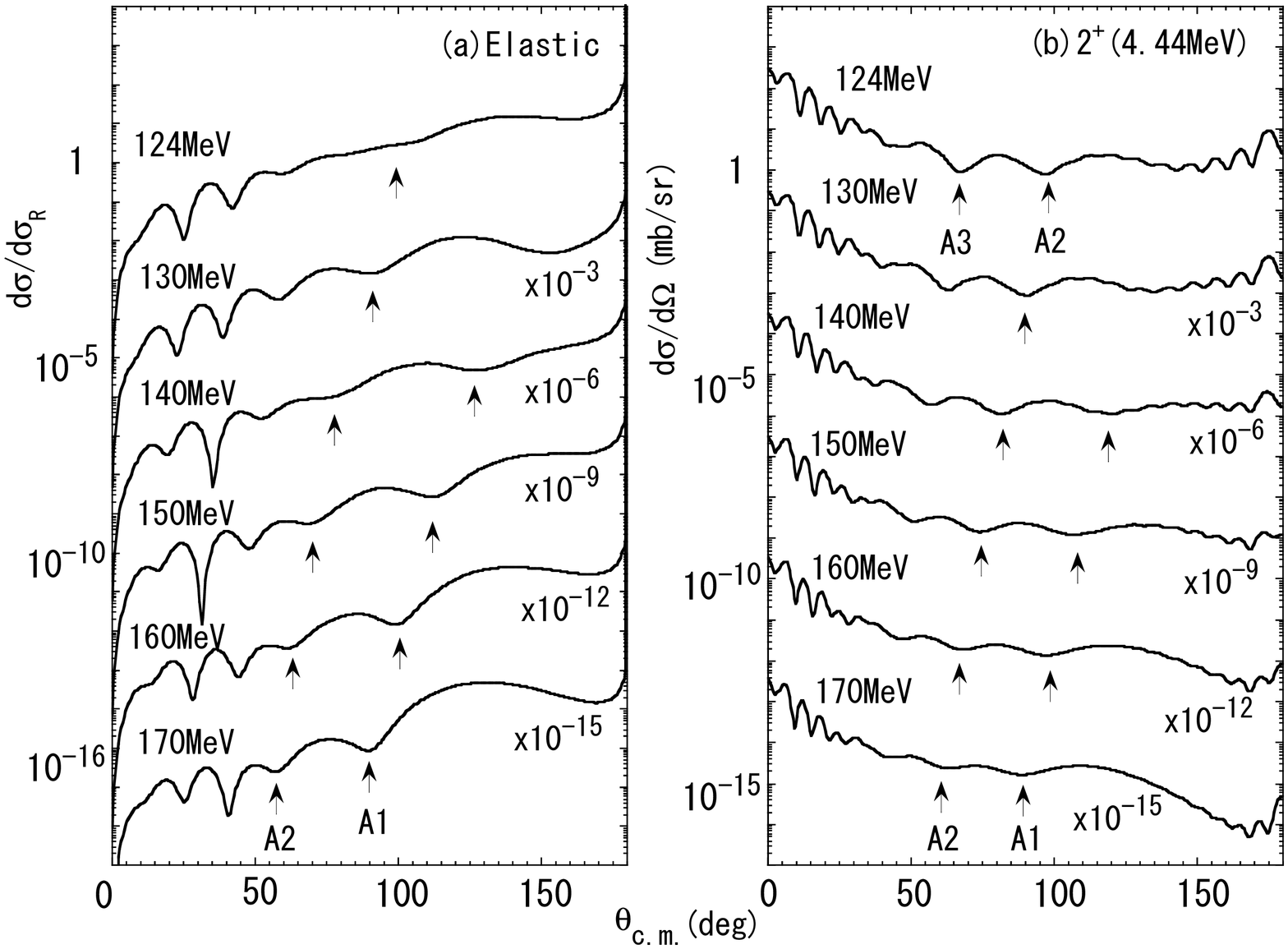}% ,angle=-90Here is how to import EPS art
\protect\caption{\label{fig.2} {
 Energy evolution of the angular distributions in   $^{16}$O+$^{12}$C  scattering calculated by switching off the imaginary potential for  $E_L$=170-124 MeV at  intervals of 10 MeV  using the coupled channels method are displayed for  (a)   elastic  (ratio to Rutherford  scattering) and (b) inelastic scattering to the $2^+$ state  of  $^{12}$C.  For the  inelastic scattering in panel (b) only the imaginary potential in the $2^+$ channel
  is switched off.  In panel (a) the  solid lines represent the farside components. The labels $A1$, $A2$ and  $A3$ indicate the position of the Airy minimum of the order one, two and three, respectively.
}
 }
\end{figure*}

\par
In Fig.~1  angular distributions of elastic and inelastic $^{16}$O+$^{12}$C  scattering at 
$E_L$=124-200 MeV,   calculated  using the coupled channels method,  
  are displayed in comparison  with  the  experimental data.
We take $N_R$=0.97  and the  imaginary potentials  given in Table I are used for all the channels.
 In Fig.~1(a) the  agreement of the calculated angular distributions for elastic scattering with the  experimental data   is comparable to  Ref.\cite{Ohkubo2014C}.
 The energy  evolution of the angles of the  Airy minimum  in the angular distributions  in elastic scattering  is consistent with that studied with the single channel optical potential  model  in the  lower energy region 62-124 MeV  in Ref.\cite{Nicoli2000,Szilner2001} and in the higher energy region 132-260 MeV  in Ref.\cite{Ogloblin2000}.   
In Fig.~1(a)  we note that for  the coupled channels  calculations in which  the imaginary potentials are switched off,    supernumerary bows with  higher order Airy minima   are  seen at  angles smaller than the $A1$ minima. The supernumerary bows  with  Airy minima up to   $A5$ (order five) for elastic scattering have been observed at lower energies between 85 MeV and 132 MeV by Szilner {\it et al.} \cite{Szilner2001}. They observed  the $A2$, $A3$ and $A4$ minima at around $\theta_{c.m.}$=94$^\circ$, 60$^\circ$  and   40$^\circ$    in the angular distribution for $E_L$=124 MeV and at around 81$^\circ$, 58$^\circ$ and 37$^\circ$ for $E_L$=132 MeV, respectively \cite{Szilner2001}. Ogloblin {\it et al.}  \cite{Ogloblin2000} observed  the $A2$  and $A3$ minima  at around 55$^\circ$ and 39$^\circ$ at $E_L$=170 MeV, respectively.  At  200  and 230 MeV,  only the $A2$ minimum was observed at around 45$^\circ$ and  35$^\circ$, respectively \cite{Ogloblin2000}.  Above this energy a   higher order Airy structure greater than $A2$  has not been  observed, although the  $A1$  appears up to 608 MeV \cite{Ogloblin2000}.
Regarding the inelastic scattering to the $2^+$ state in Fig.~1(b),  the characteristic features of the experimental angular  distributions   are reproduced well by the calculations. 
In the  angular distributions at 170 MeV calculated   by switching off only the imaginary potential in the channel of the $2^+$ state of $^{12}$C, we observe a minimum at around 60$^\circ$ faintly  in addition to the Airy minimum $A1$ at around 90$^\circ$ reported in Ref.\cite{Ohkubo2014C}. A similar minimum is  seen clearly at 124 MeV.

\begin{table}[b]
\begin{center}
\caption{ \label{Table I}
\\
The   volume integral per nucleon pair $J_V$ 
  of the DF potential  and the   imaginary potential parameters (strength $W$, radius $R$ and diffuseness $a$) used  in the coupled channels calculations are displayed.
 $J_V$ is  given only  for  the elastic  $^{16}$O(g.s.)-$^{12}$C(g.s.) channel. }
\begin{tabular}{ccccl}
 \hline
  \hline
$E_{L}$   & $J_V$       & $W$  &$R$ &$a$\\
\hline
 (MeV) &   (MeV fm$^3$) &(MeV) &(fm) &(fm)    \\    \hline
 124 &    306.1              &  16   & 5.6 & 0.30               \\
170 &    296.9               & 17   & 5.6 & 0.55             \\
181 &     294.9                &  17   & 5.6 & 0.55                \\  
  200 &   291.6                 &  18   & 5.6 & 0.60           \\
 \hline                          				   
\end{tabular}
\end{center}
\label{Table1}
\end{table}

\begin{figure}[bht]
\includegraphics[keepaspectratio,width=7.0cm] {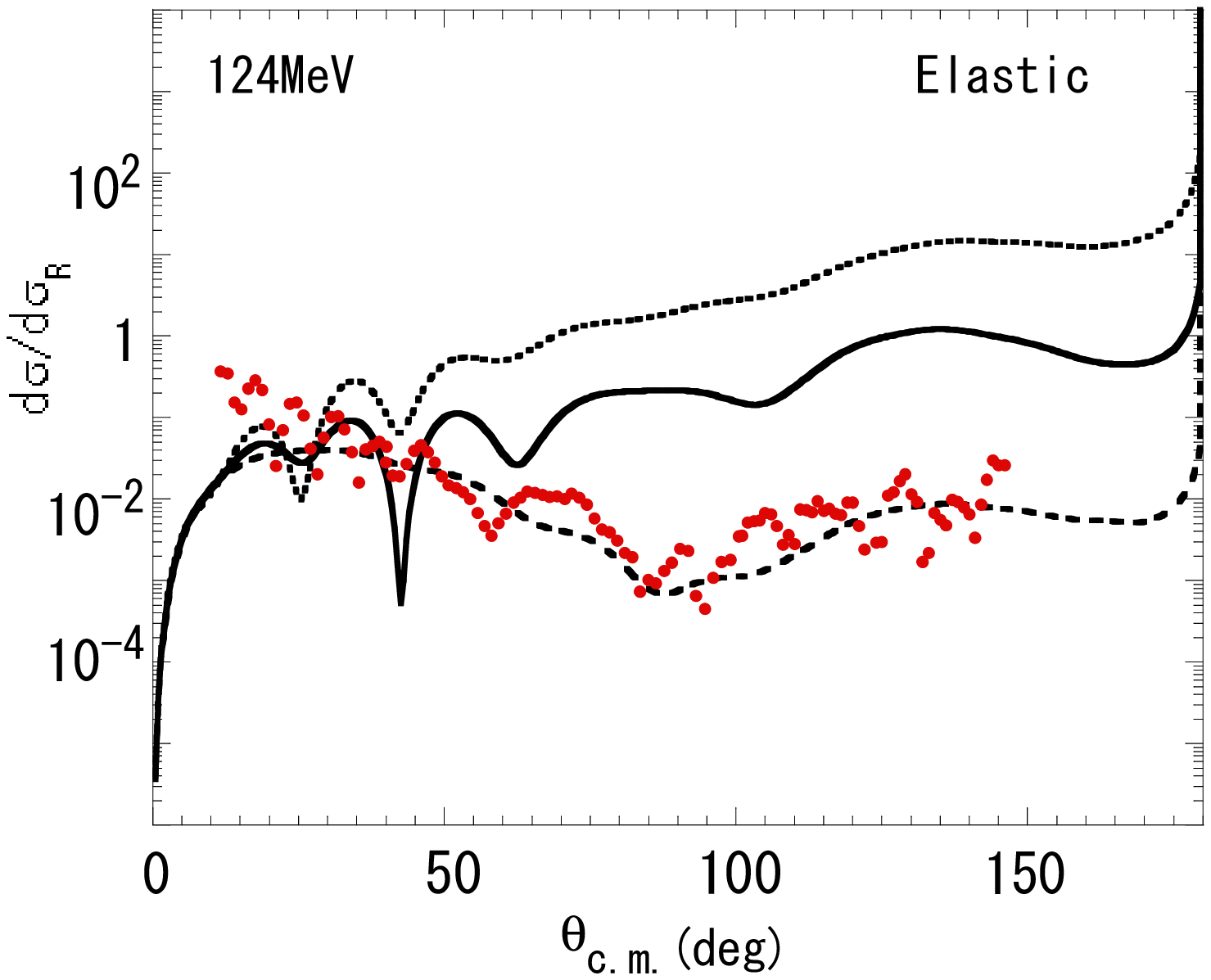}
 \protect\caption{\label{fig.3} {(Color online) 
 The farside contribution  of the angular distribution (ratio to Rutherford scattering) in $^{16}$O+$^{12}$C elastic  scattering at  $E_L$=124 MeV  calculated  using the coupled channels method  
  and  the experimental data (filled circles) \cite{Szilner2006}. 
 The dashed line, solid line and the dotted line represent the coupled channels  calculation with $W=16$ (Table I),    $W=4$ and $W=0$, respectively.
}
}
\end{figure}

\par
Since there are no inelastic scattering experimental data available between 170  and 124 MeV,
 in order to facilitate  identification of  the order of the  Airy minima at 124 MeV,
the  energy evolution  of the Airy minimum 
  between  170  and  124 MeV  is displayed   in intervals of 10 MeV in Fig.~2.  In order that the Airy minima in  Fig.~1(a) in elastic scattering can be seen clearly, in Fig.~2(a) the farside components of the cross sections (in ratio to Rutherford scattering) calculated by switching off the imaginary potential are displayed. In  Fig.~2(b) we can identify the Airy minimum $A2$ at around 63.5$^\circ$ for $E_L$=170 MeV  and 97.5$^\circ$ for $E_L$=124 MeV in the  angular distributions calculated 
 by switching off only the imaginary potential in the $^{12}$C($2^+$) channel.  
 At 124 MeV the Airy minimum   $A3$ is seen at 67.5$^\circ$. 
 The minimum at 50$^\circ$   in the angular distributions of  inelastic scattering at  200 MeV calculated 
by switching off the imaginary potentials 
in Fig.~1(b) is  found to be $A2$, which is  faintly seen  at around 50$^\circ$ in the experimental data.

 \par
In Fig.~3 the farside  contribution to angular distributions for elastic scattering at  124 MeV   is displayed for different  imaginary potential strengths. 
 The  position of the Airy minima in the coupled channel calculations does not  change for $W=0$,
4 and 16 MeV.
When absorption is switched off ($W=0$), the Airy minima appear  in the farside scattering angular distributions, $A2$, $A3$ and $A4$ at  104$^\circ$, 62$^\circ$ and 42.5$^\circ$, respectively.
By increasing absorption  to $W=4$, 
the $A2$, $A3$ and  $A4$ minima appear clearly at 103$^\circ$, 62$^\circ$ and 42$^\circ$, respectively.  
 Despite    absorption,  the supernumerary bows  in elastic scattering survive at 124 MeV.

\section{DISCUSSION}
 \par
  The mechanism   and the logic that the supernumerary bow survives in inelastic  scattering are quite similar to elastic scattering.  In fact,   the $A2$,  $A3$  and $A4$   Airy minima   at the angles around 97.5$^\circ$, 67.5$^\circ$ and    46$^\circ$    in the  angular distributions at 124 MeV calculated by  switching  off the absorption only in the $^{12}$C($2^+$) channel     are  obscured  when $W$ is increased to $W=16$. However, their angular  positions are slightly altered similar to the   elastic scattering case as shown in Fig.~1. The minimum at around 95$^\circ$
    in the experimental data of inelastic scattering, which is very close to the Airy minimum $A2$
    of the experimental data in elastic scattering, is thus assigned to be the  higher order Airy minimum $A2$.

\begin{figure}[t]
\includegraphics[keepaspectratio,width=7cm] {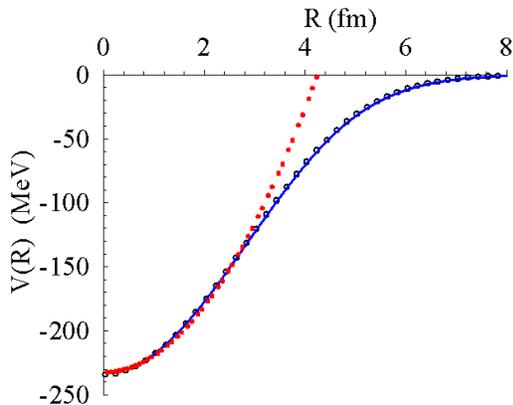}
\protect\caption{\label{fig.4} {(Color online) 
Comparison of the $^{16}$O+$^{12}$C  folding potential  for the inelastic channel to the $^{12}$C($2^+$) state (open  circles) at 124 MeV with that for elastic channel (blue solid line)    and a  Luneburg lens potential (red filled circles) with $V_0$= 233 MeV and $R_0$=4.26   fm. The  Luneburg lens potential  is given  by \cite{Michel2002} $V(R \leq R_0) = V_0 \left(  {R^2}/{R_0^2}-1 \right)$  and   $V(R > R_0) = 0$ with $R_0$ being the size of the lens. 
  }
}
\end{figure}

 \par
 The similarity of the logic of the emergence of the supernumerary Airy structure in inelastic scattering to
that in elastic scattering can be understood by looking at the real potentials that cause strong refraction and
astigmatism due to the  diffuse surface of the nuclear potential.    In Fig.~4 the folding potential for the  inelastic
$^{16}$O(g.s.)-$^{12}$C($2^+$)  channel at 124 MeV used in the coupled channels calculations in Fig.~1 is compared with that for elastic scattering and  a Luneburg lens potential. We see that the potential for the inelastic channel
(open circles) with $J_V$=303 MeVfm$^3$ is almost indistinguishable   from  the elastic channel (solid line) with $J_V$=306 MeVfm$^3$.   We also clearly see that the potential for the inelastic channel is similar to a Luneburg lens potential in the internal region.  Notch test calculations, in which the diagonal real potentials both in the elastic and 
inelastic channels are slightly modified by adding a peaked attraction located at $R=R_1$  with a small width $a_1$=0.15 fm, show that the angular distributions at angles where the Airy structure appears are sensitive to the internal region, $R_1<3$ fm, of the potential not only  in  elastic channel but also in  inelastic channel.  A notch potential  shifts the position of the Airy minimum in the angular distributions. The emergence of the nuclear rainbow with a supernumerary bow is due the properties of both the Luneburg-lens-like potential in the internal region and the diffuse attraction in the outer region of the nuclear potential  as in elastic scattering \cite{Michel2002}.

\begin{figure}[bht]

\includegraphics[keepaspectratio,width=7cm] {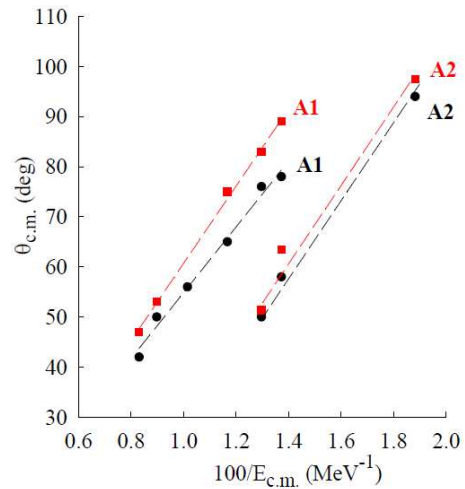}
\protect\caption{\label{fig.5} {(Color online) 
 The positions of the Airy minimum  observed  in  inelastic scattering to the $2^+$ state of  $^{12}$C  (filled squares) and elastic scattering (filled circles)  of    $^{16}$O+$^{12}$C are displayed as a function of the inverse center-of-mass  (c.m.) energies.
   The labels $A1$ and  $A2$    indicate the order of the Airy minima.
The lines are to guide the eye.}
}
\end{figure}

\par
 In Fig.~5 the energy evolution of the angular position of the Airy minimum for elastic scattering and inelastic
 scattering are displayed as a function of the inverse center-of-mass  (c.m.) energies. The Airy minimum $A1$ for elastic and inelastic scattering to the $2^+$ state of $^{12}$C at the higher energies $E_L$=260 and 281 MeV determined in Ref.\cite{Ohkubo2014C} are also included.
The Airy minimum $A1$ for inelastic scattering is slightly shifted  backward compared with that for elastic scattering. 
This backward shift is considered to be  caused by the excitation energy effect as discussed in Ref.\cite{Michel2004B}.
This backward shift  is small for $A2$. The positions of the Airy minima $A2$ for inelastic scattering vary   approximately linearly with c.m. energy similar to those for elastic  scattering \cite{Ogloblin2003}. This  seems to  support  the  assignment  of the higher order Airy minimum for inelastic scattering to the $2^+$ state of $^{12}$C. 

\par 
Finally we mention that the present model with an EDF  interaction enables us  to study   
  cluster structure with  the $^{16}$O+$^{12}$C($2^+$)  configuration  at lower  energies    and the emergence of a 
  nuclear rainbow in inelastic scattering in a unified way. This  will be reported elsewhere in a forthcoming paper. 

\section{SUMMARY}
 \par
To summarize,  we have reported the existence of a supernumerary bow in inelastic scattering 
by investigating $^{16}$O+$^{12}$C  scattering to the  $2^+$  (4.44 MeV)  state of $^{12}$C.      The systematic analysis of  rainbow scattering at $E_L$=124-200 MeV was undertaken  using an extended double folding model derived from the realistic  wave functions for $^{12}$C  and $^{16}$O calculated with a  microscopic  $\alpha$ cluster model with  a finite-range  density-dependent   nucleon-nucleon force. In the coupled channels calculations couplings to the    0$^+_1$ (0.0 MeV), $2^+$ (4.44 MeV),    3$^-$ (9.64 MeV), and  $4^+$ (14.08 MeV)  states of $^{12}$C and  the $0_1^+$ (0.0MeV),  $3^-$ (6.13 MeV) and $2^+$ (6.92 MeV) states of $^{16}$O were taken into account. The coupled channels  analysis 
made it possible to assign  the emergence of the higher order Airy minimum and the   known $A1$ minimum in the inelastic scattering cross sections to the first $2^+$ state of $^{12}$C.
 The existence of a supernumerary rainbow for the  inelastic channel of nuclear rainbow scattering in addition to the existence of a dynamically created  secondary bow \cite{Ohkubo2014} and a dynamically refracted primary bow \cite{Ohkubo2016}   for elastic scattering, which are not expected in  meteorological rainbows,  deepen the understanding  of rainbows under strong interactions.

\section{ACKOWLEDGEMENTS}
  Two of the authors (SO and YH) would like to thank the Yukawa Institute  for Theoretical Physics, Kyoto University for  the hospitality extended  during  stays in    2016 and 2017 where part of this work was done. 
 One of the authors (SO) is grateful to   Dr. Paul Suckling for careful reading of the manuscript and useful comments.
  We  thank Prof. Masayasu~Kamimura for providing the newly calculated transition densities of $^{12}$C and Prof. Florent~Haas for the tabulated form of the numerical data of Ref.\cite{Szilner2006}.


\begin{thebibliography}{aa}
\bibitem {Nussenzveig1977} 
H. M. Nussenzveig, Sci. Am. {\bf 236}, 116 (1977).
\bibitem {Airy1938} 
G. B. Airy, Trans. Camb. Phil. Soc. {\bf 6}, 379 (1838).
 
\bibitem {Jackson1999} %meteo rainbow
J. D. Jackson, Phys. Rep. {\bf 320}, 27 (1999).
\bibitem {Adam2002} 
J. A. Adam, Phys. Rep. {\bf 56},  229 (2002).
\bibitem {Hundhausen1965} 
E. Hundhausen and  H. Pauly, Z. Phys. {\bf 187}, 305 (1965).
\bibitem {Goldberg1974}
 D. A. Goldberg,
  S. M. Smith,  and G. F.   Burdzik, 
 Phys. Rev. C  {\bf 10}, 1362 (1974).
\bibitem {Khoa2007} % rainbow review 
D. T. Khoa, W. von Oertzen, H. G. Bohlen, and S. Ohkubo, 
J. Phys. {\bf G 34}, R111 (2007).
\bibitem {Brandan1997} 
M.~E.~Brandan and G.~R.~Satchler, Phys. Rep. {\bf 285}, 143 (1997)
  and references therein.
\bibitem {Michel1998}
 F. Michel, S. Ohkubo, and  G. Reidemeister, 
   Prog. Theor. Phys. Suppl. {\bf 132}, 7 (1998) 
 and references therein.
 \bibitem {Ohkubo2002} % 16O+16O=32S 
S. Ohkubo and K. Yamashita, 
 Phys. Rev.  C {\bf  66}, 021301(R) (2002).
\bibitem {Ohkubo2003} % superdeformation
S. Ohkubo, {\it Proceedings of Symposium on Nuclear Clusters: From light exotic 
 to superheavy nuclei}, 
edited by J. Jolos and W. Scheid, (EP Systema, Debrechen, Hungary
 2003) p.161; S. Ohkubo, 
Heavy Ion Physics {\bf 18},  287 (2003).
\bibitem {Nicoli1999}
M. P. Nicoli {\it et al.}, 
%, F. Haas, R. M. Freeman, N. Aissaoui, C. Beck,
%A. Elanique, R. Nouicer, A. Morsad, S. Szilner, Z. Basrak, M. E. Brandan, and
%  G. R. Satchler,
 Phys. Rev. C {\bf 60}, 064608 (1999).
\bibitem {Ogloblin1998}
A. A. Ogloblin,
Dao T. Khoa, Y. Kond\={o}, Yu. A. Glukhov, A. S. Dem'yanova,  M. V. Rozhkov, G. R. Satchler,
and S. A. Goncharov,
 Phys. Rev. C {\bf 57}, 1797 (1998).
\bibitem {Nicoli2000}
M. P. Nicoli {\it et al.}, 
Phys. Rev. C  {\bf  61},  034609 (2000).
\bibitem {Szilner2001}
S. Szilner, M. P. Nicoli,  Z. Basrak, M. Freeman, F. Haas, R. A. Morsad, 
 M. E. Brandan, and G. R. Satchler, 
 Phys. Rev. C  {\bf  64}, 064614  (2001).
 \bibitem {Ogloblin2000}
A. A. Ogloblin {\it et al.}, 
 Phys. Rev. C  {\bf 62}, 044601 (2000).
\bibitem {Ogloblin2003}
A. A. Ogloblin,  S. A. Goncharov, Yu. A. Glukhov, A. S. Dem'yanova,
 M. V. Rozhkov, V. P. Rudakov, and W. H. Trzaska,
 Phys.  At.  Nucl.  {\bf 66}, 1478 (2003).
\bibitem {Stokstad1979}
 R.~G. Stokstad, R.~M. Wieland, G.~R. Satchler, C.~B. Fulmer,
D.~C. Hensley, S.~Raman, L.~D. Rickertsen, A.~H. Snell, and P.~H.
Stelson, Phys. Rev. C {\bf 20}, 655 (1979).
\bibitem {Michel2004A}
F.~Michel and S.~Ohkubo, Eur. Phys. J.  {\bf A 19}, 333 (2004).
\bibitem {Hefter1981} %observation of supernumerary inelastic atom molecule rainbow
U. Hefter, P. L. Jones, A. Mattheus, J. Witt, K. Bergmann, and R. Schinke,
Phys. Rev. Lett. {\bf 46}, 915 (1981).
\bibitem {Gottwald1987} %Supernumerary rotational rainbows in NarHe, Ne, Ar scattering
E. Gottwald, K. Bergmann, and R. Schinke, 
J. Chem. Phys. {\bf 86}, 2685 (1987).
\bibitem{Michel2004B} F.~Michel  and S.~Ohkubo, 
Phys. Rev. C  {\bf 70}, 044609 (2004);
  Nucl. Phys. {\bf A 738}, 231 (2004).
\bibitem{Bohlen1982} % 300 MeV 12C+12C elastic rainbow and inelastic rainbow
H. G. Bohlen, 
  M. R. Clover, G. Ingold, H. Lettau, and W. von Oertzen,
Z. Phys. A {\bf 308},  121  (1982).
\bibitem{Bohlen1985}
H. G. Bohlen, X. S. Chen, J. G. Cramer, P. Fr\"{o}brich, B. Gebauer,
 H. Lettau, A. Miczaika, W. von Oertzen, R. Ulrich, and T. Wilpert,
 Z. Phys. A {\bf 322}, 241 (1985). 
\bibitem{Bohlen1993} % 350MeV 390 MeV 16O+16O inelastic 2+ 3- mixed
H. G. Bohlen,
 E. Stiliaris,  B. Gebauer,  W. von Oertzen,  M. Wilpert, Th. Wilpert, 
A. Ostrowski,  Dao T. Khoa,  A. S. Demyanova,  and A. A. Ogloblin,
Z. Phys. A {\bf 346}, 189  (1993).
\bibitem {Khoa2005}
D. T. Khoa, H. G. Bohlen, W. von Oertzen, G. Bartnitzky,
A. Blazevic, F. Nuoffer, B. Gebauer,  W. Mittig, and
P. Roussel-Chomaz,
Nucl. Phys. {\bf A 759}, 3 (2005).
\bibitem {Brandan1986}
M. E. Brandan,
A. Menchaca-Rocha, M.  Buenerd, J. Chauvin, P. DeSaintignon, G. Duharmel, D. Lebrun,
 P. Martin, G. Perrin, and J. Y. Hostachy,
Phys. Rev. C {\bf 34}, 1484 (1986).
\bibitem {Ohkubo2014C}
S. Ohkubo, Y. Hirabayashi, A. A. Ogloblin, Yu. A. Gloukhov, A. S. Dem'yanova, 
and W. H. Trzaska,
Phys. Rev. {\bf C 90}, 064617 (2014).
\bibitem{Michel2005}
 F.~Michel  and S.~Ohkubo, 
Phys. Rev. C  {\bf 72}, 054601 (2005).
\bibitem {Ohkubo2004}
 S. Ohkubo and Y. Hirabayashi,
 Phys. Rev. C {\bf 70}, 041602(R) (2004).
\bibitem {Ohkubo2007}
  S. Ohkubo and Y. Hirabayashi,
 Phys. Rev. C {\bf 75}, 044609 (2007).
\bibitem {Belyaeva2010}
T. L. Belyaeva, A. N. Danilov, A. S. Dem'yanova,  S. A. Goncharov,
 A. A. Ogloblin, and R. Perez-Torres,
Phys. Rev.  C {\bf  82},  054618 (2010).
\bibitem {Hamada2013}
  Sh. Hamada, Y. Hirabayashi, N. Burtebayev, and S. Ohkubo, 
 Phys. Rev. C {\bf 87},  024311 (2013).
\bibitem {Ohkubo2010}
 S. Ohkubo and  Y. Hirabayashi,
 Phys. Lett.  {\bf B684}, 127  (2010).
\bibitem {Hirabayashi2013}
Y. Hirabayashi and S. Ohkubo, 
Phys. Rev.  C {\bf  88}, 014314 (2013).
\bibitem {Trzaska2002}
W. H. Trzaska,
 Phys. At. Nucl.   {\bf 65}, 725 (2002).
 \bibitem {Ohkubo2014}%secondary bow
S. Ohkubo and Y. Hirabayashi, 
Phys. Rev. {\bf C 89}, 051601(R)  (2014).
\bibitem {Ohkubo2014B} %ripple
S. Ohkubo and Y. Hirabayashi, 
Phys. Rev. {\bf C 89}, 061601(R)  (2014).
\bibitem {Ohkubo2016} %dynamical primary rainbow 9Be+16O
S. Ohkubo and Y. Hirabayashi, 
Phys. Rev. {\bf C 94}, 034601  (2016).
\bibitem {Szilner2006}
S. Szilner, F. Haas, Z. Basrak, R. M. Freeman, A. Morsad, and M. P. Nicoli,
Nucl. Phys.  {\bf A 779}, 21 (2006);
S. Szilner {\it et al.}, IAEA Database EXFOR, http://www-nds.iaea.org/exfor/.
\bibitem {Okabe1995} 
S. Okabe, in {\it Tours Symposium on Nuclear Physics II}, edited
by H. Utsunomiya {\it et al.} (World Scientific, Singapore, 1995),
p. 112.
\bibitem {Suzuki1976} %OCM
Y.~Suzuki,
Prog. Theor. Phys. {\bf 55}, 1751 (1976);
 Prog. Theor. Phys. {\bf 56}, 111 (1976). 
\bibitem{Kamimura1981}
M.~Kamimura, Nucl. Phys. {\bf A 351}, 456 (1981).
\bibitem {Kobos1982}
A. M. Kobos,
 B. A. Brown, P. E. Hodgson,  G. R. Satchler, and A. Budzanowski,
 Nucl. Phys. {\bf A 384}, 65 (1982);
A. M. Kobos,
 B. A. Brown, R. Lindsaym, and G. R. Satchler,
Nucl. Phys. {\bf A 425}, 205 (1984);
 G. Bertsch, J. Borysowicz,
H. McManus, and W. G. Love, Nucl. Phys. {\bf A 284}, 399 (1977).
\bibitem{Khoa1994} 
D. T. Khoa, W. von Oertzen,  and H. G. Bohlen, 
 Phys. Rev. C  {\bf  49}, 1652 (1994).
  \bibitem {Michel2002}%Luneburg lens approach to nuclear rainbow scattering.
F. Michel, G. Reidemeister, and  S.   Ohkubo, 
Phys. Rev. Lett. {\bf 89}, 152701 (2002).
\end{thebibliography}
\end{document}